\documentclass{jaa}
\usepackage{txfonts}
\usepackage{graphicx}
\usepackage{psfig}

\usepackage[authoryear]{natbib}

\begin{document}

\title[The spectral variability of quasar SDSS J030639.57+000343.1]{The spectral variability of quasar SDSS J030639.57+000343.1}
\author[H. Guo \& M. Gu]%
       {Hengxiao Guo$^{1,2,}$
       \thanks{e-mail:hxguo@shao.ac.cn},     Minfeng Gu$^1$\\
        $^1$ Key Laboratory for Research in Galaxies and Cosmology, Shanghai Astronomical\\
         Observatory, Chinese Academy of Sciences, 80 Nandan Road Shanghai 200030, China \\
         $^2$ University of Chinese Academy of Sciences, 19A Yuquanlu, Beijing 100049, China}
   \maketitle
\label{firstpage}

\begin{abstract}
We compiled a sample of 60 quasars with spectroscopy on at least six epochs from the Sloan Digital Sky Survey (SDSS) to study the variabilities of the spectral shape, the continuum and the emission lines luminosity. In this paper, we present the results of SDSS J030639.57+000343.1. We found a strong anti-correlation between the continuum luminosity at $5100~\AA$ and the spectral index, implying a bluer-when-brighter trend. The luminosity of the broad $\rm H_\alpha$ line is proportion to the continuum luminosity at $5100~\AA$. Correspondingly, we did not find strong correlation between the equivalent width of broad $\rm H\alpha$ and the continuum luminosity, i.e. no baldwin effect of broad $\rm H\alpha$ in this source.

\end{abstract}

\begin{keywords}
Galaxies: active - galaxies: individual: SDSS J030639.57+000343.1 - techniques: spectroscopic.
\end{keywords}

\section{Introduction}

Quasars are one kind of powerful Active Galactic Nuclei (AGNs), characterized by strong and rapid variability (Schmidt 1969). There are extensive investigations on the continuum variability, especially in radio-loud AGNs (e.g., Fan et al. 1998; Ghosh 2000; Gu et al. 2011a,b), and two trends of color variation have been found. The Bluer-When-Brighter trend (BWB) was commonly found in blazars, and radio quiet AGNs as well (e.g., Wu et al. 2005; Gu et al. 2011a,b). However, the Redder-When-Brighter trend (RWB) has also been found (e.g., Gu et al. 2006; Bian et al. 2012).

From the multi-epoch spectra, the variabilities of the broad emission lines luminosity can be explored, in addition to those of the continuum luminosity and spectral shape. However, only few works have been done on the optical variabilities using multi-epoch spectroscopy. The variability of C~{\sc iv} lines has been studied for a sample of 105 quasars at two epochs (Wilhite et al. 2006). Recently, the spectral variability of FIRST bright quasars was investigated using SDSS spectra (Bian et al. 2012). We have compiled a large sample from SDSS with multi-epoch spectroscopy. As a first step, the spectral variability were explored for a sample with $\geq$ 6 epochs spectroscopy. In this paper, we present the results of SDSS J030639.57+000343.1. 

\begin{figure}
     \centering
     \includegraphics[width=0.8\textwidth]{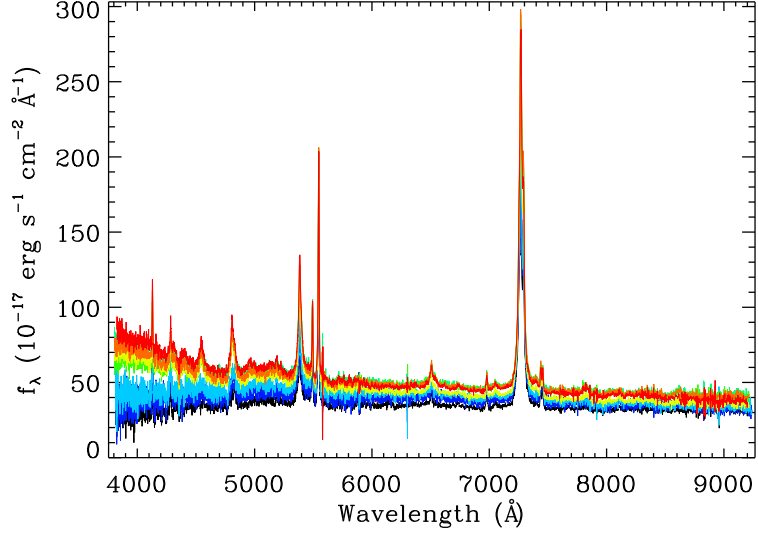}
     \caption[]{The spectra of SDSS J030639.57+000343.1 ($z=0.107$) at nine epochs in the observed frame after extinction correction.  }
\label{Fig.1}
   \end{figure}

\begin{figure}
     \centering
     \includegraphics[width=0.48\textwidth]{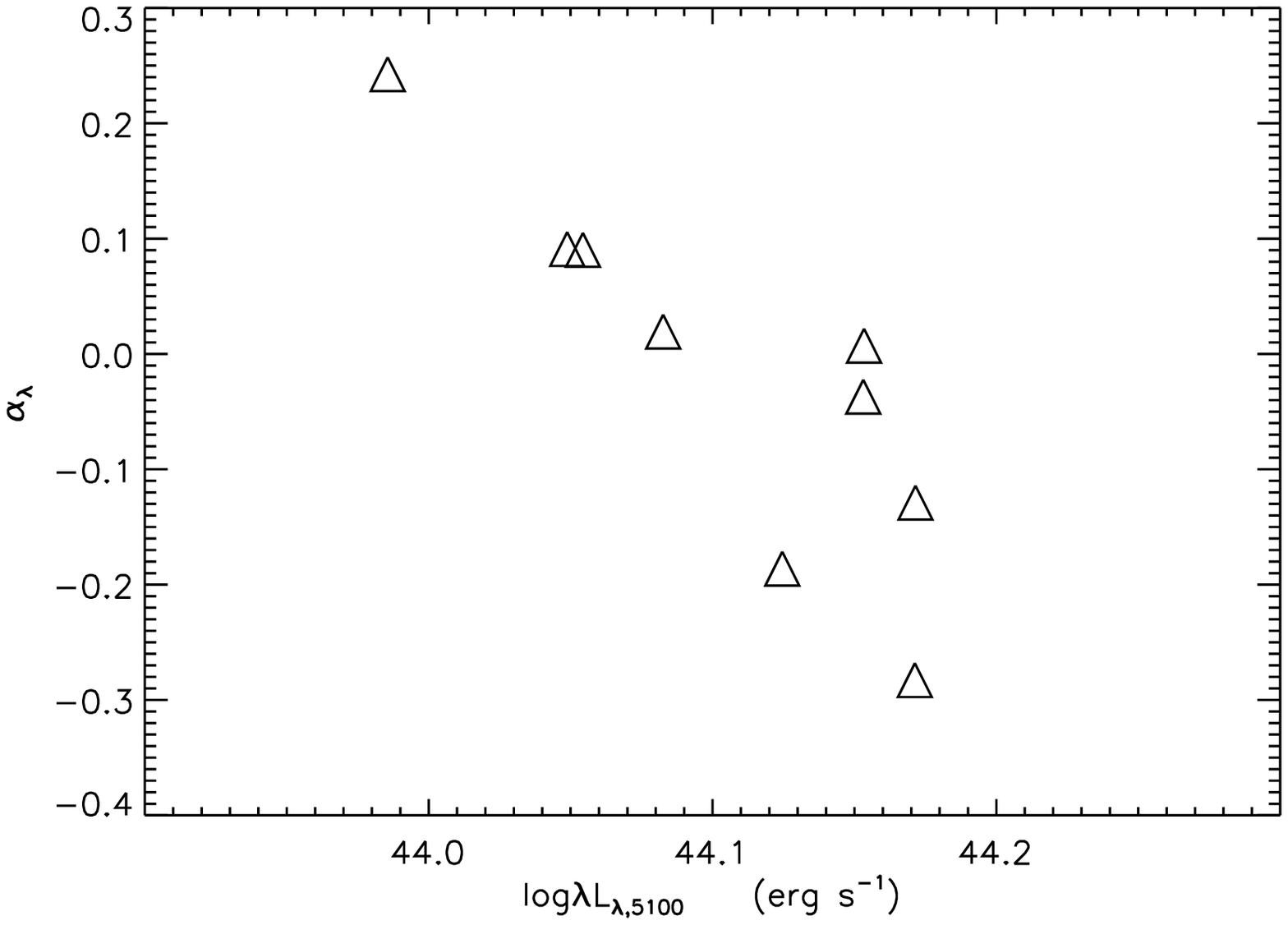}
     \includegraphics[width=0.48\textwidth]{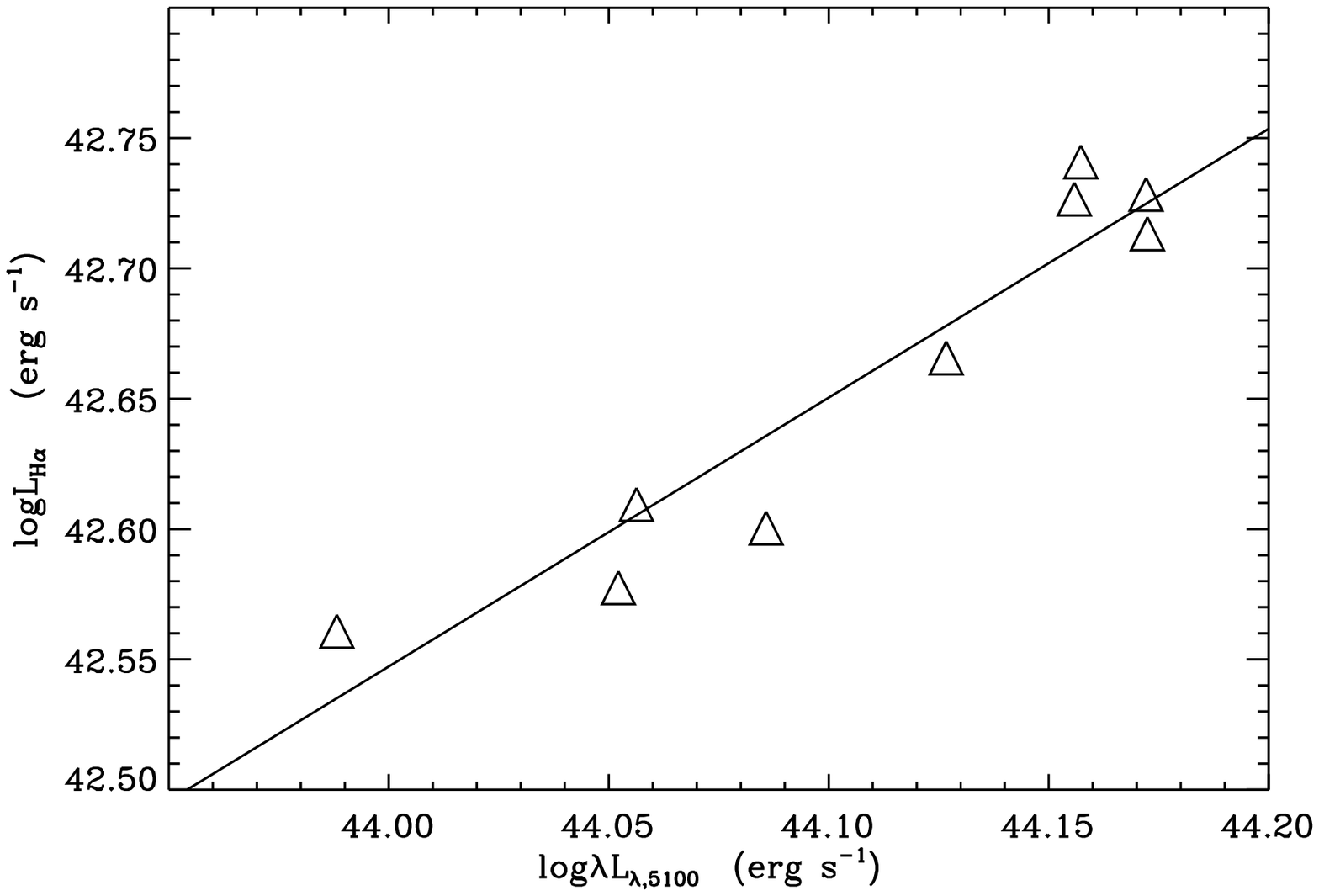}
     \caption[]{Left : the spectral index versus the continuum luminosity at 5100{\AA} at nine epochs. Right : the continuum luminosity at 5100{\AA} versus broad $\rm H_\alpha$ luminosity at nine epochs. The solid line is the linear fit. }
\label{Fig.2}
   \end{figure}

\section{Sample}

We selected 7,063 quasars having spectra on at least two epochs from 105,783 SDSS DR7 quasars in \cite{Shen11}. As a first step to study the spectral variability, we choose a subsample of 60 quasars with $\geq$ 6 epochs spectroscopy, of which the redshift ranges from 0.08 to 3.78. Following \cite{Chen09}, the spectra was firstly corrected for the galactic extinction, then transferred to the rest frame. The continuum was fitted with a power-law, and the optical and UV Fe II features and the Balmer continuum were considered.

SDSS J030639.57+000343.1 ($z=0.107$) has been observed spectroscopically nine times in SDSS. The spectra is shown in Fig. \ref{Fig.1}. We fitted two bright emission lines ($\rm H_\alpha$, $\rm H_\beta$) each with two gaussians, one for the broad and the other for narrow profiles. \cite{Shen11} estimated the black hole mass $M_{\rm bh}$ = $10^{7.51 \pm 0.13}$ $M_\odot$, and the Eddington ratio $L_{\rm bol}/L_{\rm Edd}$ = 0.24. The radio loudness $R={f_{\rm 6 cm}/ f_{\rm 2500{\AA}}}$ is given as 5.32, where $f_{\rm 6 cm}$ and $f_{\rm 2500{\AA}}$ are the flux density at 6 cm and 2500${\AA}$ at rest frame, respectively. Therefore, the source is likely a radio intermediate quasar, and the jet emission may not severely contaminate the optical continuum emission.

\section{Results}

We found a strong anti-correlation between the continuum luminosity at 5100{\AA} and spectral index $\alpha_{\lambda}$ ($f_\lambda \varpropto \lambda^{\alpha_{\lambda}}$) with the Spearman rank correlation coefficient $r=-0.85$ at 99.7$\%$ confidence level (Fig. 2, left). This implies the source is bluer when brighter, which is commonly found in our sample. A significant correlation was also found between the continuum luminosity at 5100{\AA} and broad $\rm H_\alpha$ luminosity with r = 0.867 at confidence level of 99.7$\%$ (Fig. \ref{Fig.2}, right). A linear fit shows $\rm log~\it L_{\rm H\alpha}$ = (1.03 $\pm$ 0.13)~log~($\lambda$$L_{\rm \lambda,{5100\AA}}$) - (2.83 $\pm$ 5.81). Consistent with this linear relation, we did not find significant correlation between the equivalent width of broad $\rm H_\alpha$ and the 5100{\AA} continuum luminosity, i.e. no baldwin effect of broad $\rm H_{\alpha}$ in this source. We found similar results for broad $\rm H_\beta$. The comprehensive investigations on our whole sample will be presented in a forthcoming paper (Guo \& Gu 2013, in prep.).\\

\noindent{{\textbf{Acknowledgements}}} \\
\noindent{This work is supported by the 973 Program (No. 2009CB824800), and by the NSFC grant 11073039. }

\label{lastpage}

\begin{thebibliography}{100}
\setlength{\itemsep}{0pt}
\bibitem[Bian et al.(2012)]{bian12} Bian, W.-H., Zhang, L.,Green, R., \& Hu, C.\ 2012, ApJ, {\bf759}, 88
\bibitem[Chen et al.(2009)]{Chen09} Chen, Z., Gu, M., \& Cao, X.\ 2009, MNRAS, {\bf397}, 1713
\bibitem[Fan et al.(1998)]{Fan98} Fan, J.~H., Xie, G.~Z., Lin, R.~G., \& Qin, Y.~P.\ 1998, A\&AS, {\bf133}, 217
\bibitem[Ghosh et al.(2000)]{Ghosh00} Ghosh, K.~K., Ramsey,B.~D., Sadun, A.~C., Soundararajaperumal, S., \& Wang, J.\ 2000, ApJ, 537, 638
\bibitem[Gu et al.(2006)]{Gu06} Gu, M.~F., Lee, C.-U., Pak, S., Yim, H.~S., \& Fletcher, A.~B.\ 2006, A\&A, {\bf450}, 39
\bibitem[Gu \& Ai(2011)]{2011A&A...528A..95G} Gu, M.-F., \& Ai, Y.~L.\ 2011a, A\&A, {\bf528}, A95
\bibitem[Gu \& Ai(2011)]{Gu11} Gu, M.~F., \& Ai, Y.~L.\ 2011b, A\&A, {\bf534}, A59
\bibitem[Guo \& Gu et al.(2013)]{Guo13} Guo, H.~X., Gu, M.~F., in preparation
\bibitem[Schmidt(1969)]{Schmidt69}Schmidt, M.\ 1969,Contemporary Physics, Volume {\bf1}, 467
\bibitem[Shen et al.(2011)]{Shen11} Shen, Y., Richards, G.~T.,Strauss, M.~A., et al.\ 2011, ApJS, {\bf194}, 45
\bibitem[Wilhite et al.(2006)]{Wilhite06} Wilhite, B.~C., VandenBerk, D.~E., Brunner, R.~J., \& Brinkmann, J.~V.\ 2006, ApJ, {\bf641}, 78
\bibitem[Wu et al.(2005)]{Wu05} Wu, J., Peng, B., Zhou, X.,et al.\ 2005, AJ, {\bf129}, 1818
\end{thebibliography}
\end{document}